\documentclass[twocolumn,showpacs,preprintnumbers,amsmath,amssymb, superscriptaddress]{revtex4}

\usepackage{amsmath}
\usepackage{amssymb}
\usepackage{amsfonts}
\usepackage{fancyhdr}
\usepackage{fancybox}
\usepackage{ulem}

\usepackage{graphicx}
\usepackage{float}

\usepackage{amsthm}

\DeclareMathOperator{\Tr}{Tr}

\newcommand{\paren}[1]{{\left( #1 \right)}}

\newcommand{\var}[2]{{\delta{#1}\over\delta{#2}}}

\newcommand{\Vtr}[3]
{\paren{\begin{array}{c} #1 \\ #2 \\ #3 \end{array}}}

\makeatletter
\def\loweq@align#1#2{\lower.6ex\vbox{\baselineskip\z@skip\lineskip\z@
    \ialign{$\m@th#1\hfil##\hfil$\crcr#2\crcr=\crcr}}}
\def\lowsim@align#1#2{\lower.6ex\vbox{\baselineskip\z@skip\lineskip\z@
    \ialign{$\m@th#1\hfil##\hfil$\crcr#2\crcr\sim\crcr}}}
\def\geqq{\mathrel{\mathpalette\loweq@align >}}
\def\leqq{\mathrel{\mathpalette\loweq@align <}}
\def\grsim{\mathrel{\mathpalette\lowsim@align >}}
\def\lesssim{\mathrel{\mathpalette\lowsim@align <}}
\def\gsim{\mathrel{\mathpalette\lowsim@align >}}
\def\lsim{\mathrel{\mathpalette\lowsim@align <}}
\makeatother
\newcommand{\grless} 
{ {\, \raise-.24em\hbox{$<$} \hspace{-0.8em} \raise.31em\hbox{$>$}\, } }
\newcommand{\lessgr} 
{ {\, \raise-.24em\hbox{$>$} \hspace{-0.8em} \raise.31em\hbox{$<$}\, } }
\newfont{\bg}{cmr10 scaled\magstep4}                    
\newcommand{\bigzerou}{\smash{\lower1.7ex\hbox{\bg 0}}}

\newcommand{\nn}{\nonumber \\ }

\newcommand{\C}{{\mathbb C}}

\newcommand{\HH}{{\cal H}}

\newcommand{\crl}[1]{[-\infty,\infty]}

\newcommand{\ket}[1]{|{#1}\rangle}
\newcommand{\kb}[1]{|{#1}\rangle\!\langle{#1}|}
\newcommand{\bk}[1]{\langle{#1}|{#1}\rangle}

\newcommand{\bra}[1]{\langle{#1}|}
\newcommand{\bkt}[2]{\langle{#1}|{#2}\rangle}
\newcommand{\Ref}[1]{(\ref{#1})}

\newcommand{\wt}{\widetilde}

\newcommand{\av}[1]{\langle#1\rangle}
\newcommand{\cc}{{\rm c.c.}}
\newcommand{\wtU}{U}
\newcommand{\vv}[1]{{\boldsymbol{#1}}}

\begin{document}
\title{Quantum Brachistochrone}
  \author{Alberto Carlini}
 \email{carlini@th.phys.titech.ac.jp}
 \affiliation{Department of Physics, Tokyo Institute of
 Technology, Tokyo, Japan}
\author{Akio Hosoya}
 \email{ahosoya@th.phys.titech.ac.jp}
 \affiliation{Department of Physics, Tokyo Institute of
 Technology, Tokyo, Japan}
\author{Tatsuhiko Koike}
 \email{koike@phys.keio.ac.jp}
 \affiliation{Department of Physics, Keio University, Yokohama, Japan}
\author{Yosuke Okudaira}
 \email{okudaira@th.phys.titech.ac.jp}
 \affiliation{Department of Physics, Tokyo Institute of
 Technology, Tokyo, Japan}

\date{November 4, 2005}

\begin{abstract}
We present a general framework for finding the time-optimal evolution and
the optimal Hamiltonian for a quantum system with a given set of initial and 
final states.  
Our formulation is based on the variational principle and is analogous 
to that for the brachistochrone in classical mechanics. 
We reduce the problem to a formal equation for the Hamiltonian which depends on 
certain constraint functions specifying the range of available Hamiltonians.  
For some simple examples of the constraints, we explicitly find the optimal solutions.
\end{abstract}

\pacs{03.67.-a, 03.67.Lx, 03.65.Ca, 02.30.Xx, 02.30.Yy}

\maketitle

In quantum mechanics one can change a given 
state to another by applying a suitable Hamiltonian on the system. 
In many situations, e.g. quantum computation, it is desirable to know 
the pathway in the shortest time. 

In this paper we consider the problem of finding the time-optimal path for the 
evolution of a pure quantum state and the optimal driving Hamiltonian. 
Recently, a growing number of works related to this topic have appeared. 
For instance, Alvarez and G\'omez  \cite{alvarez} showed that the quantum state in Grover's algorithm \cite{grover},
known as the optimal quantum search algorithm \cite{zalka},
actually follows a geodesic curve derived from the Fubini-Study metric
in the projective space.
Khaneja {\it et al.} \cite{khaneja} and Zhang {\it et al.} \cite{zhang}, using a
Cartan decomposition scheme for unitary operations, discussed the time 
optimal way to realize a two-qubit universal unitary gate under the condition 
that one-qubit operations can be performed in an arbitrarily short time.
On the other hand, Tanimura {\it et al.} \cite{tanimura} gave an adiabatic solution to the 
optimal control problem in holonomic quantum computation, in which a 
desired unitary gate is generated as the holonomy corresponding to the minimal 
length loop in the space of control parameters for the Hamiltonian.
Schulte-Herbr\"uggen {\it et al.} \cite{schulte} exploited the differential geometry of the projective 
unitary group
to give the tightest known upper bounds on the actual time complexity of some basic modules 
of quantum algorithms.
More recently, Nielsen \cite{nielsen} introduced a lower bound on the size of 
the quantum circuit necessary to realize a given unitary operator based on the geodesic 
distance, with a suitable metric, between the unitary and the 
identity operators.
However, a general method for generating the time optimal Hamiltonian evolution
of quantum states was not known until now.

Here we are going to study this problem by exploiting the analogy with the so-called brachistochrone problem in classical mechanics and the elementary properties of quantum mechanics.
In ordinary quantum mechanics the initial state and the Hamiltonian of a physical system are given and one has to find the final state using the Schr\"odinger equation.
In our work we generalize this framework so as to optimize a certain cost functional with 
respect to the Hamiltonian as well as the quantum states.  
The cost functional quantifies the efficiency to get the target state from a given initial state and
depends on the physical situation.
In this paper we focus our attention on the time optimality but it is straightforward to generalize our methods to other cost functions. 

In the brachistochrone problem one has to find the shape of a friction-free tube connecting two points 
and with a particle running inside subject only to homogeneous gravity.
The solution, a segment of a cycloid, can be found using the variational principle for 
the evolution time $T[\vv{x}(t)]=\int \frac{ds}{v}$,
where the parameter $s$ specifies the length of the tube from the initial to the current position 
$\vv{x}(t)$ of the particle, i.e. $ds^2=|d\vv{x}|^2$. 
The magnitude of the particle's velocity is $v  :=\frac{ds}{dt}=\sqrt{2(E-V(\vv{x}))/m}$,
where $E$ is the conserved energy and $V$ is the gravitational potential.

Let us now move to the time optimization problem in the quantum case.
We want to minimize the total amount of time necessary for changing 
a given initial state $\ket{\psi_i}$ (belonging to an $n$-dimensional Hilbert
space $\HH$) to a given final state $\ket{\psi_f}$, by suitable choice of a 
(possibly time-dependent) Hamiltonian $H(t)$. 
In our problem the quantum state $\ket{\psi(t)}$ and the Hamiltonian $H(t)$ 
are the dynamical variables, and the action is defined as
\begin{align}
  \label{eq-action}
  S(\psi,H,\phi,\lambda)=&\int dt\bigg[
    \frac{\sqrt{\bra{\dot\psi}(1-P)\ket{\dot\psi} } } {\Delta E}
  \nn
  &\qquad +\left(i\bkt{\dot\phi}{\psi}
      +\bra\phi H\ket\psi+\cc \right)
  \nn
  &\qquad +{\lambda}\paren{\Tr {\wt H{}^2}/{2}-\omega^2} \biggr]. 
\end{align}
Here $P(t):=\kb{\psi(t)}$ is the projection to the state $\ket{\psi(t)}$, 
$\wt H:=H- (\Tr H)/n$ is the traceless part of the Hamiltonian, 
$(\Delta E)^2:=\bra\psi H^2\ket\psi-\bra\psi H\ket\psi{}^2$ is the
energy variance and $\omega$ is a given nonzero constant
(an overdot denotes differentiation with respect to the time $t$). 
We have chosen units in which Planck's constant $\hbar$ is equal to one.

The first term in the action \Ref{eq-action} gives the time duration 
for the evolution of $\ket{\psi(t)}$,  expressed in terms of 
the Fubini-Study line element $ds^2=\bra{d\psi}(1-P)\ket{d\psi}$ 
on the projective space $\C P^{n-1}$.
The second term guarantees, through the Lagrange multiplier $\ket{\phi(t)}\in\HH$, that $\ket{\psi(t)}$ and $H(t)$ satisfy the Schr\"odinger equation and that the squared norm 
$\bk\psi =1$ is conserved.
The third term, through the Lagrange multiplier $\lambda$, generates a constraint for the Hamiltonian. 
Such a constraint is necessary because otherwise one would be able to find a path with
arbitrarily small time duration just by rescaling the Hamiltonian as $H\mapsto \alpha H$, with $\alpha >1$, to make the energy fluctuations $\Delta E$ large.
This corresponds to the fact that physically only a finite amount of resources (e.g., a finite magnetic 
field) is available. 
Here we consider a typical example which we call the isotropic constraint (we will consider more general constraints later). 
The constraint is imposed on $\wt H$ rather than $H$, since
the difference between the highest and the lowest energy levels in $H$, and not the value of the energy levels themselves, is important for the physical system. 
The problem should be mathematically formulated on the projective space $\C P^{n-1}$ rather than on $\HH$, because the overall phase of the state $\ket\psi$ is of no significance in quantum mechanics.
In fact, the action \Ref{eq-action} is invariant under the $U(1)$ gauge transformation 
$ (\ket{\psi},H,\ket\phi,\lambda)
  \mapsto 
  (e^{-i\theta}\ket{\psi}, 
  H+\dot\theta, 
  e^{-i\theta}\ket{\phi}, 
  \lambda)$, where $\theta(t)$ is a real function. 
Note that the Hamiltonian $H$ plays the role of the gauge potential 
(there is also another symmetry, $\ket\phi\mapsto\ket\phi+i\gamma\ket\psi$,
where $\gamma$ is a real constant). 

Let us now derive the equations of motion. The variation of \Ref{eq-action}
with respect to $\bra\phi$ leads to the Schr\"odinger equation 
\begin{align}
  i\ket{\dot\psi}=H\ket\psi. 
  \label{eq-Sch}
\end{align}
In particular, this implies $\bra{\dot\psi}(1-P)\ket{\dot\psi}=(\Delta E)^2$, or
\begin{align}
  \label{eq-AA-rel}
  ds=\Delta E~dt,
\end{align}
which was found by Aharonov and Anandan \cite{aa} and leads to a rigorous
formulation of the time-energy uncertainty principle.
The variation with respect to $\lambda$ gives the constraint 
$ \Tr {\wt H}^2=2\omega^2$. 
The variation with respect to $\bra\psi$, upon using \Ref{eq-Sch},
yields 
\begin{align}
  \label{eq-dpsi}
  i\paren{\frac{H-\av H}{2(\Delta E)^2}}^{\text{\large$\cdot$}} \ket\psi
  -i\ket{\dot\phi}+H\ket\phi=0. 
\end{align}
Finally, the variation with respect to  $H$, after use of \Ref{eq-Sch}, implies 
\begin{align}
 \frac{\{H, P\}-2\av HP}{2(\Delta E)^2}-\lambda \wt H
 -(\ket\psi\bra\phi +\ket\phi\bra\psi)=0
  \label{eq-dH}
\end{align}
where a bracket $\av\bullet$ denotes the expectation value with
respect to $\ket\psi$ and $\{A, B\}=AB+BA$. 
Equation  \Ref{eq-dpsi} and the trace of \Ref{eq-dH} imply that  $\bkt{\psi}{\phi}$ is a purely imaginary constant.
Then,  the expectation value of \Ref{eq-dH} gives  $\av{\wt H}=0$, which is equivalent to $\av H=(\Tr H)/n$,  or $\wt H=H-\av H$. 
Applying \Ref{eq-dH} to $\ket\psi$, we have 
\begin{align}
  \ket\phi=
  \left [\paren{\frac1{2(\Delta E)^2}-\lambda}\wt H+\bkt{\psi}{\phi}\right ]\ket\psi, 
  \label{eq-lambda}
\end{align}
and inserting \Ref{eq-lambda} back into \Ref{eq-dH}, we obtain 
\begin{align}
  \label{eq-H}
  \wt H=\wt HP+P\wt H. 
\end{align}
Furthermore, the energy variance is constant, i.e.
$(\Delta E)^2=\av{\wt H{}^2}=\Tr {\wt H}^2/2=\omega^2$.
Substituting \Ref{eq-lambda} into \Ref{eq-dpsi}, we have 
$(\lambda \wt H)^{\text{\large$\cdot$}}\ket\psi=0$
which, after multiplication by $\bra\psi H$, implies that 
$\lambda$ is constant. 
We then obtain
\begin{align}
  \label{eq-dotH}
  \dot {\wt H}\ket{\wt\psi}=0, 
\end{align}
where we have introduced $\ket{\wt\psi}:=\exp[i\int_0^tdt\av H]\ket{\psi}$.
In conclusion, the equations to be solved have reduced to \Ref{eq-H} 
and \Ref{eq-dotH}.
Equation \Ref{eq-H} gives an expression for the optimal Hamiltonian and
\Ref{eq-dotH} gives the optimal time evolution of the quantum state.
In fact, the state $\ket{\wt\psi}$ satisfies the Schr\"odinger equation with Hamiltonian $\wt H$, and \Ref{eq-H} implies
\begin{align}
\label{wtH}
  \wt H=i(\ket{\dot{\wt\psi}}\bra{\wt\psi}-\ket{\wt\psi}\bra{\dot{\wt\psi}}). 
\end{align}
The derivative $\ket{\dot{\wt\psi}}$ is
orthogonal to $\ket{\wt\psi}$ because $\av {\wt H}=0$. 
Therefore, equation \Ref{eq-dotH} reads
\begin{align}
  \label{eq-geod}
  (1-\wt P)\ket{\ddot{\wt\psi}}=0,
\end{align}
where $\wt P=\kb{\wt\psi}=P$.
This is the geodesic equation for the
Fubini-Study metric on $\C P^{n-1}$, which is suggested by the observation that 
the first term in the action \Ref{eq-action} becomes $\int ds$ for constant $\Delta E $. 
We also easily see that $\dot{\wt H}=0$ from \Ref{wtH} and \Ref{eq-geod}.

One can solve equation \Ref{eq-geod} using $(\Delta E)^2=\omega^2$, finding: 
\begin{align}
\label{geod-sol}
  \ket{\wt\psi (t)}=
 \cos\omega t ~\ket{\wt\psi(0)}
  +\frac{\sin\omega t}{\omega} ~\ket{\dot{\wt\psi}(0)}.
\end{align}
It is then easy to rewrite \Ref{geod-sol} and $\wt H$ in terms of the Gram-Schmidt 
orthonormalized initial state $\ket{\psi_i}$ and final state $\ket{\psi_f'}$ as
\begin{align}
\label{grover1}
  \ket{\wt\psi (t)}&=
  \cos\omega t\ket{\psi_i}
  +\sin\omega t\ket{\psi_f'},
  \\
  \label{grover2}
  \wt H&=i\omega(\ket{\psi_f'}\bra{\psi_i}
  -\ket{\psi_i}\bra{\psi_f'}). 
\end{align}
As a result, the whole Hamiltonian is given by $H(t)=\wt H+\av{H(t)}$, 
where $\av {H(t)}$  is an arbitrary real function corresponding to the gauge degree 
of freedom. 
The optimal time is $T=\frac{1}{|\omega |}\arccos |\bkt{\psi_f}{\psi_i}|$.

Let us now generalize the quantum brachistochrone problem to the case of a more
general set of $m$ constraints for the Hamiltonian $H(t)$. 
This extension is physically relevant since, for example, interactions between more than
two qubits are not controllable in practical cases.
Instead of \Ref{eq-action} we consider the action 
\begin{align}
  \label{eq-gen-action}
  &S(\psi,H,\lambda,\mu)=\int dt\bigg[
    \frac{\sqrt{\bra{\dot\psi}(1-P)\ket{\dot\psi} } } {\Delta E}
  &
  \nn
  &+\left(i\bkt{\dot\phi}{\psi}
      +\bra\phi H\ket\psi+\cc \right)
  +\sum_{a=1}^m \lambda^a f_a(H) \biggr], 
  &
\end{align}
where the $f_a$  ($a=1, ... m$) are functions mapping a Hermitian
operator into a real number.
As we have already discussed, the most natural case to consider is when the $f_a$ 
are actually functions of $\wt H$, but the argument below is valid also for arbitrary $f_a$.
Among the equations of motion, 
the constraint now generalizes to $f_a(H)=0$,
while \Ref{eq-H} and \Ref{eq-dotH} become
\begin{align}
  &F=FP+PF,
  \label{eq-gen-F}
  \\
  &\paren{\dot{F}+i[\wt H, F]}\ket\psi=0,
  \label{eq-gen-muF}
\end{align}
with the operator
\begin{align}
F(H):= \sum_a\lambda^a\paren{\var {f_a}H-\av{\var {f_a}H}P}.
\end{align}
In particular, equation \Ref{eq-gen-F} guarantees that  $\Tr F=\av F=0$. 
Note that, for $m=1$ and $f=\Tr \wt H{}^2/2 -\omega^2$, equations \Ref{eq-gen-F} and 
\Ref{eq-gen-muF} reproduce \Ref{eq-H} and \Ref{eq-dotH} of the isotropic case.

We can formally integrate \Ref{eq-gen-F} and \Ref{eq-gen-muF} to obtain 
\begin{align}
  \label{eq-gen-FG}
 &F=\wtU{}F(0)\wtU^\dagger,
\end{align}
where $F(0)$ is a constant Hermitian operator which satisfies $F(0)=\{F(0), P(0)\}$ 
and the unitary operator $U$ is given as a functional of $\wt H$, 
\begin{align}
\label{eq-gen-U}
U[\wt H](t):= {\hat T} e^{-i\int_0^t\wt H dt},
\end{align}
with $\hat T$ the time-ordered product.
Thus, given the constraints $f_a(\wt H)=0$, one can explicitly write the left hand side 
of \Ref{eq-gen-FG} as  a function of $\wt H$ and solve \Ref{eq-gen-FG} to obtain 
the optimal Hamiltonian $\wt H$.
This is our main result.

As an explicit example of this general framework we may consider the
following case of one qubit subject to the two constraints 
\begin{align}
  f_1(H):&=\Tr {\wt H{}^2}/{2}-\omega^2=0,
\label{eq-ex-constr1}
\\
  f_2(H):&=\Tr(\wt H\sigma_z)=0,  
  \label{eq-ex-constr2}
\end{align}
where $\sigma_j$ ($j=x,y,z$) are the Pauli matrices. 
This corresponds to the physical situation where one can apply 
a magnetic field with $x$ and $y$ components only. 
By \Ref{eq-gen-FG} and $\Tr F=0$, we have 
\begin{align}
  \label{eq-ex-F}
  F=\lambda_1\wt H+\lambda_2\sigma_z=UF(0)U^\dagger.   
\end{align}
Eliminating $\wt H$ in \Ref{eq-ex-F} via $i\dot U=\wt H U$ and solving \Ref{eq-ex-F}, we get 
$U=\exp [i\chi\sigma_z]\exp[-iF(0)\int_0^t dt/\lambda_1]$, 
where $\chi(t):=\int_0^t(\lambda_2/\lambda_1)dt$. 
From \Ref{eq-ex-constr1}-\Ref{eq-ex-F} and the above formula for $U$, we find
that the $\lambda_j$ are constants and we can simplify $U$ and $\wt H$ as
\begin{align} 
U&=e^{i\Omega t\sigma_z}e^{-i[\wt H(0)+\Omega\sigma_z]t}, 
\label{eq-ex-U1}
\\
  \wt H&=e^{i\Omega t\sigma_z}\wt H(0)e^{-i\Omega t\sigma_z},
\label{eq-ex-H}
\end{align}
where $\Omega :=\lambda_2/\lambda_1$.  

Let us consider the case in which the initial state is on the equator
of the Bloch sphere $\C P^1$, $\av{\sigma_z}=0$. 
Without loss of generality, we may choose $P(0)=(1+\sigma_x)/2$,
i.e. the spin initially points towards the positive $x$ direction. 
Then the condition $F(0)=\{F(0), P(0)\}$ and the constraints \Ref{eq-ex-constr1} and \Ref{eq-ex-constr2} 
imply that $\wt H(0)=-\omega\sigma_y$,
and from \Ref{eq-ex-U1} and \Ref{eq-ex-H} we finally obtain
\begin{align}
  \av{{\boldsymbol\sigma}} (t)&=\Vtr
  {  \cos 2\Omega t \cos 2\Omega' t +\frac{\Omega}{\Omega'} \sin 2\Omega t \sin 2\Omega' t }
  { - \sin 2\Omega t \cos 2\Omega' t +\frac{\Omega}{\Omega'} \cos 2\Omega t \sin 2\Omega' t }
 { \frac{\omega}{\Omega'}\sin 2\Omega' t}
 \label{eq-ex-sol1}
  \\
 \wt H(t)&= -{\boldsymbol \sigma}\cdot {\bf B}(t)~~~;~~~{\bf B}(t)=\omega\Vtr
 {\sin 2\Omega t}
 {\cos 2\Omega t}
 {0}, 
 \label{eq-ex-sol2}
 \end{align}
 where $\Omega':=\sqrt{\omega^2+\Omega^2}$.
 We can interpret ${\bf B}(t)$ in \Ref{eq-ex-sol2} as a magnetic field rotating with
 angular velocity $2\Omega$.
Equation \Ref{eq-ex-sol1} is not a geodesic on $\C P^1$ unless $\Omega=0$, 
in which case the orbit is a great circle in the $xz$-plane. 
The energy fluctuation (which is also the speed of the state)
is now time-dependent, $\Delta E(t)=|\omega|\left [1-\left (\frac{\Omega}{\Omega'} \sin 2\Omega' t\right )^2 \right ]^{1/2}$.
We notice that the constraint \Ref{eq-ex-constr2} in general reduces the speed of the state.
 
If the final state $\ket{\psi_f}$ is given, the angular velocity $2\Omega$ 
and the time duration $T$ are also determined. 
For instance, let us assume $P(T)=(1-\sigma_x)/2$, 
which is the antipodal point of $P(0)$ in $\C P^1$. 
Then we have the conditions $2 |\Omega |T=k\pi$ and $2\Omega' T=l\pi$, 
where $k$ and $l$ are integers such that $l>k\geq 0$ and $k+l$ is odd. 
Thus we get
\begin{align}
 \label{eq-ex-gT}
  |\omega |T=\frac{\pi}{2}\sqrt{l^2-k^2}, 
  \quad \left |\frac{\Omega}{\omega}\right |=\frac{k}{\sqrt{l^2-k^2}}. 
\end{align}
\begin{figure}[H]
 \begin{center}
  \resizebox{5cm}{!}{\includegraphics{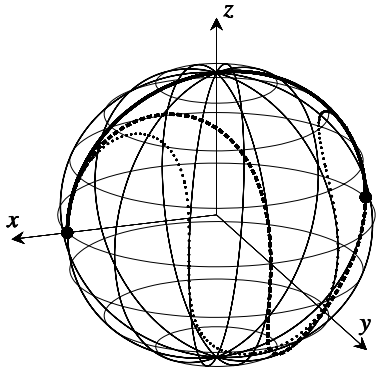}}
  \caption{Locally optimal curves on the Bloch sphere.
The solid, thick dotted and thin dotted curves correspond, respectively, to the solutions
with $|\omega | T= \frac{\pi}{2}, \frac{\pi}{2}\sqrt3$ and $\frac{\pi}{2}\sqrt5$.
They have zero, one and two nodes along the great circle on the $xy$-plane, respectively.}
 \end{center}
\end{figure}
The concrete solutions for \Ref{eq-ex-gT}, in ascending order of $T$, are
$(|\omega |T, \left |\frac{\Omega}{\omega}\right |) =
  \paren{\frac{\pi}{2}, 0},
  \paren{\frac{\pi}{2}\sqrt{3}, \frac{1}{\sqrt{3}}}, 
  \paren{\frac{\pi}{2}\sqrt{5}, \frac{2}{\sqrt{5}}}, \cdots$. 
These solutions respectively have zero, one, two $\cdots$ nodes along the great circle
$\av{\sigma_z}=0$. 
The first one is a geodesic, while the others are not. 
On the other hand, if the final state is not in the $xz$-plane, 
there is no geodesic solution.
At this stage we have to caution the reader that solutions may be only locally optimal,
and we have to find out the globally optimal one by comparing time durations $T$.
In the above case, the first curve without nodes is globally optimal.
The second curve becomes globally optimal if the final state is its first node.
The solutions are depicted in Fig. 1.

In summary,  in analogy to the classical brachistochrone problem we have formulated a variational principle to find the optimal Hamiltonian and the optimal quantum state evolution, for given initial and final states and a set of available Hamiltonians. 
As a particular application of our methods, one might want 
to first evaluate the optimal Hamiltonian, e.g. by means of a classical
computer, and then perform the fastest possible quantum experiment or quantum computation.
As a future development of the present research we do not see any obstacle to generalize our formulation to mixed states.
The relation to gate complexity in the standard paradigm of quantum computation remains to be investigated, though our point of view is that the time complexity 
(see, e.g., Schulte-Herbr\"uggen {\it et al.} \cite{schulte} and references therein) is more physical or even more practical. 
For example, equations \Ref{grover1} and \Ref{grover2} essentially solve Grover' s search problem 
\cite{alvarez, grover}.
However, to show the square-root speed up with respect to the classical case one still has 
to identify the computational step corresponding to a single oracle call.
We hope that the present variational approach to the time optimality problem in quantum mechanics opens up  novel systematic investigations of optimal quantum computation.

%
This research was partially supported by the Ministry of Education,
Science, Sports and Culture of Japan, under grant No. 09640341 (A.H. and T.K.), and 
by the COE21 project on `Nano-scale quantum physics' at Tokyo Institute of Technology
(A.C., A.H. and Y.O.).

\bibliographystyle{alpha}
\bibliography{abbrv,textbooks,QuantumComputation,stallisEntropy}

\end{document}